\documentclass[twocolumn, aps, prl, amsmath, superscriptaddress]{revtex4-1}
\usepackage{graphicx}
\usepackage{epsfig}
\usepackage[cp1251]{inputenc}
\usepackage[english]{babel}
\usepackage{amsmath}
\usepackage{amssymb}
\usepackage{amstext}
\usepackage{color}
\usepackage[normalem]{ulem}

\begin{document}

\title{Generation of a localized microwave magnetic field by coherent phonons in a ferromagnetic nanograting}
\author{A. S. Salasyuk}
\affiliation{Ioffe Institute, RAS, Politechnycheskaya 26, 194021 St. Petersburg, Russia.}
\author{A. V. Rudkovskaya}
\affiliation{Ioffe Institute, RAS, Politechnycheskaya 26, 194021 St. Petersburg, Russia.}
\author{A. P. Danilov}
\affiliation{Experimentelle Physik 2, Technische Universit\"{a}t Dortmund, Otto-Hahn-Str 4a, 44227 Dortmund, Germany.}
\author{B. A. Glavin}
\affiliation{Department of Theoretical Physics, V.E. Lashkaryov Institute of Semiconductor Physics, 03028 Kyiv, Ukraine. }
\author{S. M.~Kukhtaruk}
\affiliation{Experimentelle Physik 2, Technische Universit\"{a}t Dortmund, Otto-Hahn-Str 4a, 44227 Dortmund, Germany.}
\affiliation{Department of Theoretical Physics, V.E. Lashkaryov Institute of Semiconductor Physics, 03028 Kyiv, Ukraine. }
\author{M. Wang}
\affiliation{School of Physics and Astronomy, University of Nottingham, Nottingham NG7 2RD, United Kingdom.  }
\author{A. W. Rushforth}
\affiliation{School of Physics and Astronomy, University of Nottingham, Nottingham NG7 2RD, United Kingdom.  }
\author{P. A. Nekludova}
\affiliation{Institute of Nanotechnology of Microelectronics, RAS, Leninsky Prospekt, 32A, 119991 Moscow, Russia.}
\author{S. V. Sokolov}
\affiliation{Institute of Nanotechnology of Microelectronics, RAS, Leninsky Prospekt, 32A, 119991 Moscow, Russia.}
\author{A. A. Elistratov}
\affiliation{N. L. Dukhov All-Russia Research Institute of Automatics, 127055 Moscow, Russia.}
\author{D. R. Yakovlev}
\affiliation{Ioffe Institute, RAS, Politechnycheskaya 26, 194021 St. Petersburg, Russia.}
\affiliation{Experimentelle Physik 2, Technische Universit\"{a}t Dortmund, Otto-Hahn-Str 4a, 44227 Dortmund, Germany.}
\author{M. Bayer}
\affiliation{Ioffe Institute, RAS, Politechnycheskaya 26, 194021 St. Petersburg, Russia.}
\affiliation{Experimentelle Physik 2, Technische Universit\"{a}t Dortmund, Otto-Hahn-Str 4a, 44227 Dortmund, Germany.}
\author{A. V. Akimov}
\affiliation{School of Physics and Astronomy, University of Nottingham, Nottingham NG7 2RD, United Kingdom.  }
\author{A. V. Scherbakov}
\affiliation{Ioffe Institute, RAS, Politechnycheskaya 26, 194021 St. Petersburg, Russia.}
\affiliation{Experimentelle Physik 2, Technische Universit\"{a}t Dortmund, Otto-Hahn-Str 4a, 44227 Dortmund, Germany.}

\begin{abstract}
{A high-amplitude microwave magnetic field localized at the nanoscale is a desirable tool for various applications within the rapidly developing field of nanomagnetism. Here, we drive magnetization precession by coherent phonons in a metal ferromagnetic nanograting and generate ac-magnetic induction with extremely high amplitude (up to $10$~mT) and nanometer scale localization in the grating grooves. We trigger the magnetization by a laser pulse which excites localized surface acoustic waves. The developed technique has prospective uses in several areas of research and technology, including spatially resolved access to spin states for quantum technologies.}
\end{abstract}

\maketitle

\noindent The exploration of magnetism at the nanoscale continues to be a rapidly developing field. Magnetic recording with ultrahigh densities~\cite{Varvaro_2016} for data storage, magnetic resonant imaging with nanometer resolution~\cite{Degen_2009, Mamin_2013} for medicine and biology, addressing the magnetic states of atoms~\cite{Dutt_2007, Neumann_2008, Wu_2010, Schuster_2010, Kubo_2011} for quantum computing, and ultrasensitive magnetic sensing~\cite{Maze_2008} are the most prominent examples within the multifaceted research field of nanomagnetism. Most of the proposed concepts and prototypes utilize oscillating (ac-) magnetic fields with frequencies from millions up to hundreds of billions of cycles per second (10${}^{6}$-10${}^{11}$ Hz). The oscillating magnetic fields are used to override the coercivity of ferromagnetic grains~\cite{Thirion_2003}, to set atomic magnetic moments to a desired state ~\cite{Degen_2009, Mamin_2013, Maze_2008}, and to encode quantum information into spin states~\cite{Dutt_2007, Neumann_2008, Wu_2010, Schuster_2010, Kubo_2011, Reilly_2015}. These examples utilize conventional methods for the generation of ac-magnetic fields: an external rf-generator in combination with a microwire~\cite{Degen_2009, Mamin_2013, Dutt_2007, Neumann_2008, Maze_2008, Thirion_2003, Reilly_2015} or a microwave cavity~\cite{Wu_2010, Schuster_2010, Kubo_2011, Reilly_2015}. This methodology cannot be applied at the nanometer scale. A key breakthrough would be nanoscale generation of high-amplitude, monochromatic ac-magnetic fields. This would open the possibility to address neighboring nano-objects, e.g. spin qubits, independently, and to reduce the energy consumption in magnetic devices. It is however a challenging task to reach this goal because current technologies do not allow one to control the frequency, bandwidth and amplitude of an ac-magnetic field on the nanoscale.
\begin{figure}
\begin{center}
\includegraphics[width=8.6cm,
keepaspectratio]{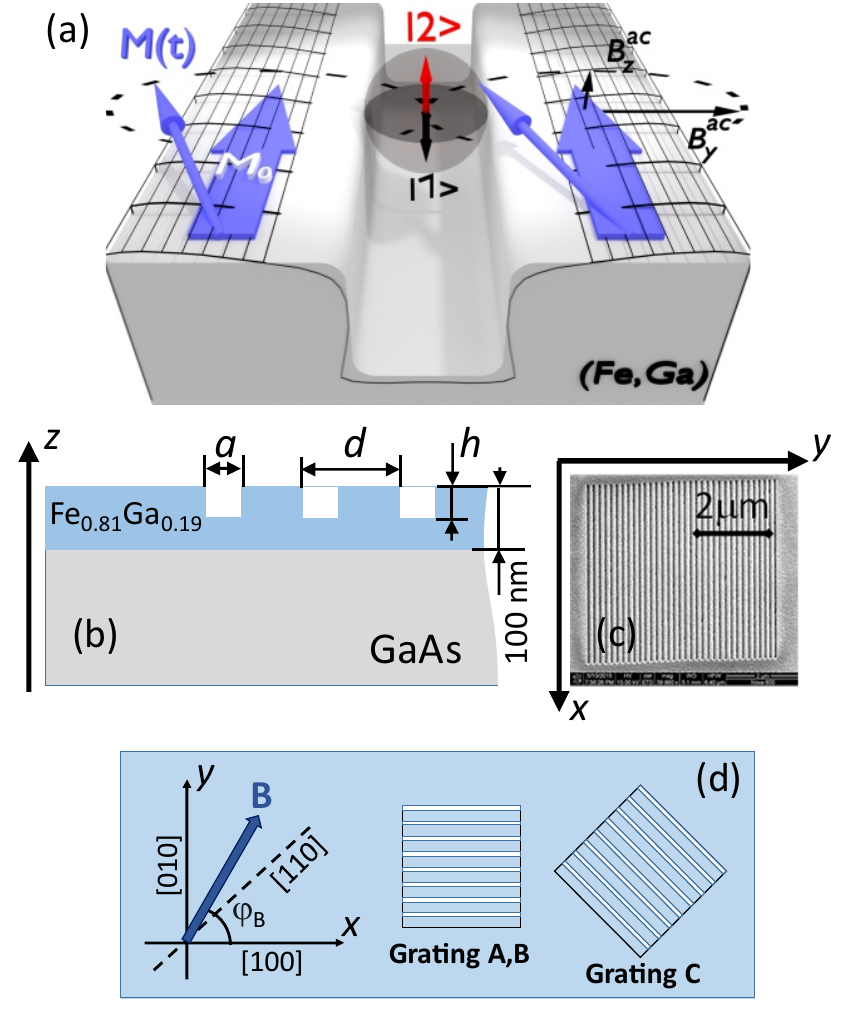}
\end{center}
\caption{(a) Schematic figure to illustrate how surface acoustic waves excited by a fs-pulsed laser drive the magnetization precession in a ferromagnetic grating. The strongly localized high amplitude ac-magnetic field can control the state of a spin qubit located inside the groove (schematically shown as sphere). (b) Schematic figure of the grating; (c) SEM image of Grating A and (d) orientations of Gratings A, B and C.
}\label{Fig1}
\end{figure}

An efficient way to generate a high-frequency ac magnetic field is to induce coherent magnetization precession in a ferromagnet. The magnetization of ferromagnetic metals may be as large as $2$ T. Precessional motion with frequencies of $10$ GHz allows the generation of high-amplitude microwave magnetic fields on the picosecond time scale. The magnetization precession can be driven by dc- spin polarized currents~\cite{Kiselev_2003}. This approach is realized in microwave generators based on spin torque nano-oscillators, but has severe limitations, e.g. in combining large amplitudes and high frequencies~\cite{Chen_2016}. Coherent phonons, bulk~\cite{Scherbakov_2010, Kim_2012} or surface~\cite{Yahagi_2014, Janusonis_2015} acoustic waves, have been successfully used for exciting the magnetization precession in ferromagnetic films.  The effect of a surface acoustic wave (SAW) on the magnetic order in a ferromagnetic structure can be sufficiently strong to switch the magnetization at very low energy cost~\cite{Ref_1,Ref_2,Ref_3}. Thus, the coherent magnetization precession resonantly driven by a monochromatic SAW may achieve large amplitudes and provide a strong ac- magnetic induction. Generation of acoustic waves with a narrow band in the frequency range of magnetization precession may be achieved by fabricating structures hosting localized phonon (nanomechanical) modes, e.g. ferromagnetic phonon cavities ~\cite{Jager_2015}. In such structures the coherent phonons drive the magnetization precession for extended times achieving ac-magnetic field quality factors $>$100~\cite{Jager_2015}. However, in unpatterned ferromagnetic films the ac-magnetic field remains localized inside the ferromagnetic material due to demagnetization and cannot be utilized in the environment, even at nanometer distances. The recipe for the transformation of magnetization precession to magnetic induction in free space is to fabricate a patterned ferromagnetic film as we report in the present work.

The aim of this Letter is to introduce a \textit{new paradigm for nanomagnetism} by fabricating a grating on a nanometer thick ferromagnetic layer for the generation of an ac-magnetic field in free space on the nanometer scale with an amplitude $\sim10$~mT and a quality factor $\textstyle{>100}$. We excite the nanostructure with femtosecond optical pulses in order to generate  SAWs, which drive the magnetization precession, as shown schematically in Fig. 1(a). The stripes and grooves in the metal ferromagnetic film form a lateral grating. At equilibrium, the magnetization $\textstyle{{\bf M}_0}$ lies along a direction determined by the magnetic anisotropy and the external magnetic field \textbf{B}. In the absence of the external magnetic field, the magnetization \textbf{M${}_{0}$} lies preferably along the stripes. The coherent magnetization precession excited by the optical pulses generates an ac-magnetic field at a frequency \textit{f}${}_{M}$ which depends on \textbf{B }and has a value of $\sim$10 GHz. The ac- induction \textbf{B${}^{ac}$ }is determined by the precessional magnetization amplitude $\textstyle{\Delta {\bf M}}$ and is highly localized in the groove. The SAWs excited by the same optical pulses in the metal grating~\cite{Bonello_2001} contribute to excitation of the magnetization precession~\cite{Yahagi_2014, Janusonis_2015} and drive it during the phonon lifetime in the grating. The frequencies of the excited SAWs, \textit{f}${}_{ph}$, are set by the grating parameters and the elastic properties of the ferromagnetic film and substrate. At resonance, when $\textstyle{f_{ph}=f_M}$, we expect maximum amplitude of the ac-magnetic field.

For the practical realization we use a grating fabricated from a ferromagnetic metallic alloy of iron and gallium (Galfenol). Galfenol has a saturation magnetization \textit{M}${}_{0}$=1.8 T and a large magnetostriction~\cite{Atulasimha_2011}. These properties make Galfenol a perfect candidate for strain-controlled microwave spintronics~\cite{Parkes_2013}.

We fabricated the gratings in a $\rm 100\!-\!nm$ thick $\rm Fe_{81}Ga_{19}$ film, deposited by magnetron sputtering on a $\rm (100)\!-\!GaAs$ substrate. The film is crystalline, and possesses a bulk-like magnetostriction and cubic magnetocrystalline anisotropy with orientation of the crystallographic axis along those of the GaAs substrate~\cite{Parkes_2013}. Three gratings with a design shown in Fig.~1(b) were fabricated in the film by focused ion beam milling. They had a lateral size of $\rm 5\times5 \,\mu m^2$, a groove width $\textstyle{a=40}$~nm and a lateral period $\textstyle{d=150}$~nm. We studied gratings with different depths and orientations relative to the in-plane crystallographic axes. The orientation of the structures relative to the crystallographic directions and the chosen coordinate system are shown in the bottom panel of Fig~1(d). We focus mainly on the structure with grooves of depth $\textstyle{h=40}$~nm oriented along the [100] crystallographic direction (Grating A). Figure 1(c) shows the scanning electron microscopy image of this grating. The second structure, Grating B, has the same orientation, but shallower grooves, $\textstyle{h=8}$~nm. The third structure, Grating C, has deep grooves, $\textstyle{h=40}$~nm, but is oriented along the $\textstyle{[1\bar{1}0]}$-direction, thus, rotated in the film plane by 45${}^\circ$ relative to Grating A.

We modeled the optical excitation of SAWs in the gratings using the COMSOL Multiphysics$\textstyle{^\circledR}$ software~\cite{Comsol} using the elastic parameters from Refs.~\cite{Clark_2012, Madelung_2004}. We focus on the low-frequency modes which can be brought into resonance with the magnetization precession and find their spectrum to be strongly dependent on the grating orientation and depth. In Grating A several modes are optically excited simultaneously, but only one of them, at frequency $\textstyle{{f}_{ph}}$$\mathrm{\approx}$13~GHz has a superior quality factor $>$ 100. In Grating B with \textit{h}=8~nm there are two localized SAW modes with $\textstyle{{f}_{ph}}$$\mathrm{\approx}$14 and 16.5~GHz, which are optically excited. They have similar quality factors $>$ 100, but considerably smaller amplitudes than in Grating A for the same excitation density. Grating C has two spectrally closely lying long-living modes around $\textstyle{{f}_{ph}}$=19~GHz. The large difference in \textit{f}${}_{ph}$ with changing the grating orientation (Gratings A and C) originates from the strong elastic anisotropy in Galfenol~\cite{Clark_2012}. Detailed information about the optically excited SAW modes may be found in the Supplemental Material~\cite{SM}.

The external magnetic field, \textbf{B} is used to tune the precession frequency $\textstyle{f_M}$ across the phonon frequencies $\textstyle{f_{ph}}$. To describe the in-plane orientation of \textbf{B}, we introduce the angle  ${\varphi}_{B}$ relative to the x-axis ([100]-direction) in accordance with Fig. 1(d). To monitor the magnetization precession, we use a magneto-optical pump-probe technique. Both pump and probe were split from the same optical regenerative amplifier (wavelength 1030~nm, pulse duration 200~fs, pulse repetition rate 5~kHz). The pump pulse focused to a spot with $\textstyle{\rm 50\,\mu m}$ diameter excites  the grating area with an energy density of 10 mJ/cm${}^{2}$. The linearly polarized probe pulse is focused to a spot with $\textstyle{\rm 1.5\,\mu m}$ diameter. The detected signals of transient Kerr rotation reflect the time evolution of the normal component of the precessing magnetization, $\textstyle{\Delta M_z(t)}$~\cite{Hiebert_1997}.

Figure~2 summarizes the main experimental results for the Grating A. The upper curves in Fig.~2(a) and (b) show the typical precessional signal and its fast Fourier transform (FFT), detected in a non-patterned film area. There, the optically excited precession decays with a characteristic time $\textstyle{\tau=0.1}$~ns, independent of the direction and strength of the applied magnetic field~\cite{Kats_2016}. The corresponding FFT spectrum consists of a single line with 8-GHz spectral width. Its spectral position shifts to higher frequencies with the increase of \textit{B}~\cite{Kats_2016}.
\begin{figure}
\begin{center}
\includegraphics[width=8.3cm,
keepaspectratio]{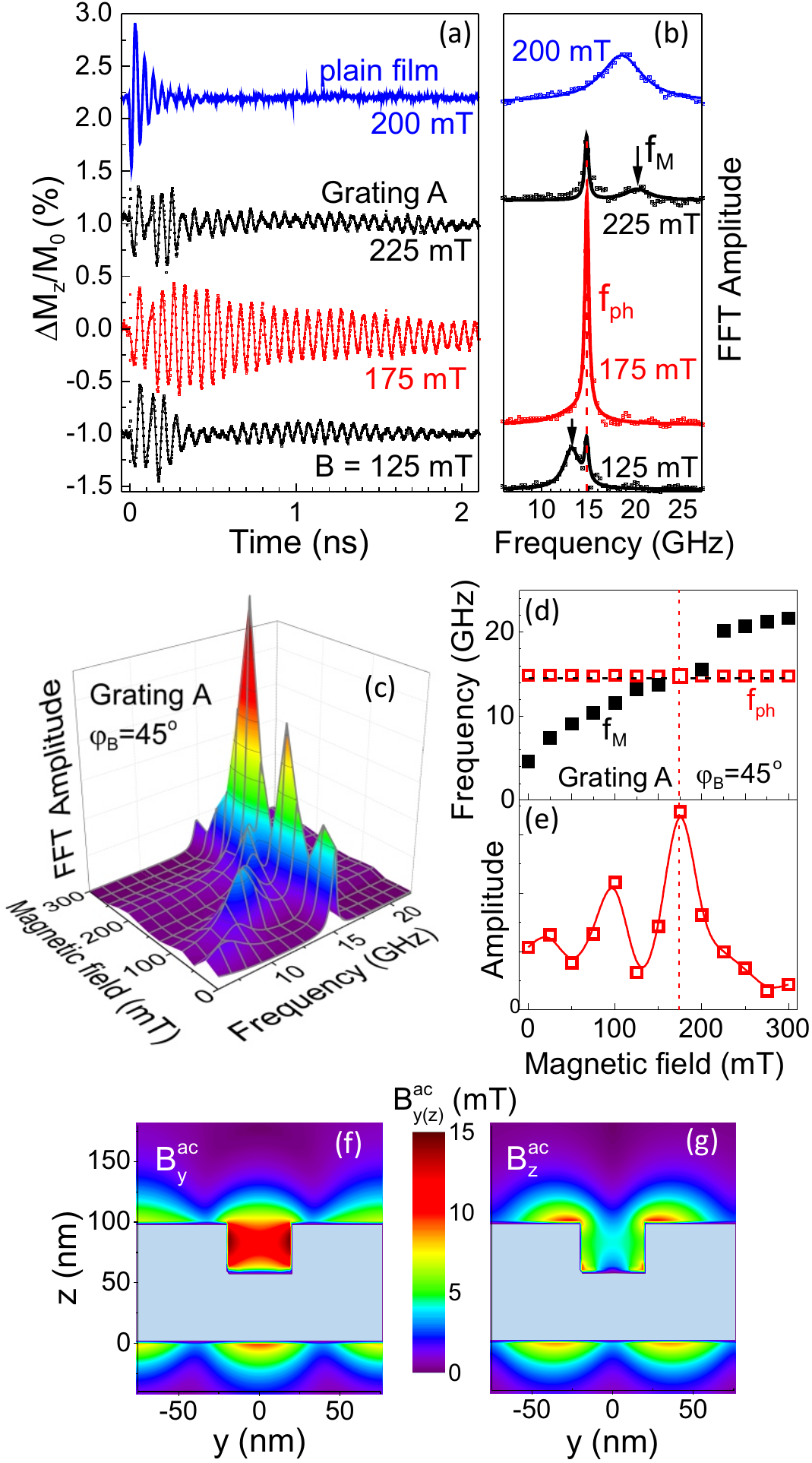}
\end{center}
\caption{Temporal evolution of the magnetization signals (a) and their FFTs (b) measured in Grating A for various applied fields $B$. The red dashed line in (b) indicates the frequency of the localized SAW mode. (c) 3D plot showing the frequency-field evolution of the precession spectra in Grating A. (d) Field dependencies of the spectral positions of the two lines in the precession spectra of Grating A.  (e) Spectral amplitude at $\textstyle{f_{ph}=15}$~GHz vs $B$. (f) and (g) Calculated spatial distributions for the $\textstyle{y}$ and $\textstyle{z}$ projections of the ac-magnetic field generated by the SAW-driven magnetization in Grating A.
}\label{Fig2}
\end{figure}

The signal detected in Grating A is significantly different from the case of the plain Galfenol film. The three lower curves in Fig. 2(a) show $\textstyle{\Delta M_z(t)}$ measured for three values of the external magnetic field applied at $\varphi_{B}$=45${}^{\circ}$. The magnetization precession lasts much longer compared to the non-patterned film. We observe oscillations over a time interval longer than 3 ns. In the FFT spectra [Fig. 2(b)] of these signals, we see a narrow line with spectral width of about 0.4 GHz (lower limit set by the duration of the monitored time range) centered at \textit{f}$\mathrm{\approx}$15 GHz. Its spectral position does not change with \textit{B}. The frequency \textit{f}=15 GHz is close to the calculated frequency \textit{f}${}_{ph}$ of the localized SAW mode in this structure. This leads us to the conclusion that the long-living magnetization precession results from a long-living SAW in the patterned film. The 2 GHz difference between the calculated (\textit{f}${}_{ph}$) and experimentally measured values may be due to some uncertainty of the elastic parameters of Galfenol, which depend sensitively on the alloy composition and temperature~\cite{Clark_2012}. There is a second line visible in the spectra at \textit{B}=125 and 225 mT. This line is broader than the one at \textit{f}=15 GHz and its spectral position depends on \textit{B}. We attribute this line to the rapidly decaying free precession excited by the optical pump pulse. The amplitude of the precession signal, i.e. the maximum value of $\textstyle{\Delta M_z(t)}$, depends on magnetic field. At \textit{B}=175 mT we reach the resonance condition $\textstyle{f_{ph}=f_M}$ where $\textstyle{\Delta M_z(t)}$ has maximum amplitude. Its FFT spectrum shows only one spectral line with 0.4~GHz width. This is the \textit{main experimental result} of the present work and we explain it by a magneto-phonon resonance in the studied ferromagnetic grating.

Figure~2(c) shows the frequency-field map of the precession in Grating A, where the evolution of the two spectral lines with changing field is presented in more detail. One line gradually shifts to higher frequencies with increasing \textit{B} while the spectral position of the other line at \textit{f}=15 GHz is field-independent. Figures 2(d) and (e) show the field dependences of the spectral positions of these two lines and the spectral amplitude of a Lorentzian fit to the resonance in the FFT spectra at \textit{f}=15 GHz, respectively. At \textit{B}=175 mT,\textit{ }corresponding to the crossing of the two frequencies, only one line appears in the spectrum and its amplitude reaches maximum value. In addition to the main maximum at \textit{B=}175 mT in Fig.~2(e), there are two more maxima at lower fields of 25 and 100 mT. We attribute these maxima to resonances of the long-living SAW with high-order magnon states that are quantized along the \textit{z}-axis~\cite{Damon_1961}. The related details are beyond the scope of the present paper.

The ac-magnetic induction generated by the precessing magnetization exists also outside of the Galfenol film. Figures 2 (f) and (g) show the calculated spatial distributions of the in-plane and out-of-plane amplitudes, $B_{y}^{ac} $and $B_{z}^{ac} $, respectively, generated by the precessing \textbf{M} at the resonant condition (\textit{B}=175 mT). The spatial distribution of \textbf{B${}^{ac}$} is determined by the film pattern, and also by the nonuniform profile of the long-living SAW mode. For the detailed description of the model used in the calculations see the Supplemental Material~\cite{SM}). \textbf{B${}^{ac}$} is highly localized within $\sim$10 nm and for our pump excitation density reaches an amplitude up to 10 mT at 15 GHz frequency. Moreover, $B_{y}^{ac} $ and $B_{z}^{ac} $ oscillate coherently giving rise to an elliptical polarization induction. In the Supplementary Video~\cite{SM} this is illustrated together with the oscillation pattern of the SAW strain. We want to emphasize that the model calculations based on the structural and anisotropy parameters of the grating are in good agreement with the experimental data. The calculated precession amplitude $\textstyle{\Delta M_z/M_0}$ averaged over the $\textstyle{\rm 1.5\!-\!\mu m}$ area of detection is $0.9\%$, which is close to the measured value.
\begin{figure}
\begin{center}
\includegraphics[width=8.6cm,
keepaspectratio]{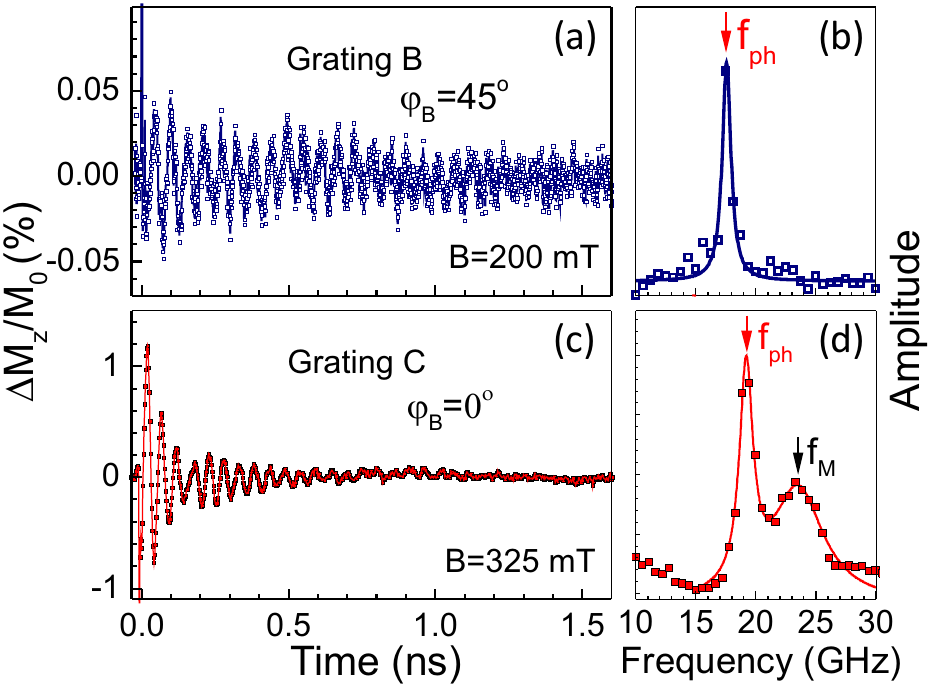}
\end{center}
\caption{Results for Gratings B (a,b) and C (c,d). Panels (a) and (c) show temporal precession signals and their FFTs are shown in panels (b) and (d). Red and black vertical arrows indicate frequencies of localized SAW and ferromagnetic resonance, respectively.
}\label{Fig3}
\end{figure}

For Grating B and C we also observe resonant driving of the magnetization, but the efficiency is much smaller than in Grating A.  Figures 3(a) and (b) show the signal $\textstyle{\Delta M_z(t)}$ and its FFT spectrum, respectively, measured in Grating B at $\varphi_{B}$=45${}^{\circ}$ and \textit{B}=200 mT. The long-living 17~GHz precession and the corresponding narrow spectral line are clearly seen, but the amplitude is smaller than in Grating~A. The small amplitude of the ac-magnetic field in the shallow grooves is consistent with the calculated frequency and amplitude for the localized SAW (see~\cite{SM}). The results for Grating C, which is rotated by 45${}^{\circ}$ relative to grating A, are shown in Fig. 3(c) and (d). The lower efficiency and faster decay of the precession compared to Grating A are due to the elastic anisotropy and the correspondingly smaller torque generated by the SAW.

The ultrafast heating and dynamical strain generated by the optical pulse modify the magnetic anisotropy and induce the precessional response of the magnetization~\cite{Kats_2016}. Here, for analysis of the long-living precession driven by the localized SAW we concentrate only on the modulation of the magnetic anisotropy by the components \textit{u}${}_{ij}$ (\textit{i},\textit{j} = \textit{x},\textit{y},\textit{z},) of the dynamical strain tensor, namely the torque produced by the strain components of the SAW on the magnetization~\cite{SM}. We consider only cubic magneto-elastic terms and assume that the magnetic field is strong enough resulting in close to parallel alignment of $\textstyle{{\bf M_0}}$ and $\textstyle{{\bf B}}$ (see~\cite{SM} for details). Then, in Grating A with grooves orientated along [100], the torque density produced by the SAW has only a \textit{z} component
\begin{equation} \label{Eq1}
\displaystyle{Q_z=-b_1 M_0u_{yy} \sin\!{(2 \varphi_B)},}
\end{equation}
where $\textstyle{b_1}$ is the magneto-elastic coefficient. We see that the magnetization precession is driven only by the strain component \textit{u${}_{yy}$ }and the driving efficiency depends on the external magnetic field direction. $\textstyle{Q_z}$ is maximum at $\varphi_{B}$=45${}^{\circ}$ and zero at $\varphi_{B}$=0 and 90${}^{\circ}$, which agrees with our experimental observations. Further, it is easy to show that in Grating C no strain components in the SAW contribute to driving the magnetization precession. Experimentally, the $\textstyle{\Delta M_z(t)}$ detected in Grating C (Fig. 3(c) and (d)) shows a finite SAW-induced precession, but its contribution to the detected signal is much weaker than in structure A (compare the signals in Grating A and C at $\textstyle{t>0.5}$~ns). The residual driving effect may be due to an in-plane uniaxial component of the magnetic anisotropy of the Galfenol film~\cite{Jager_2013} and the related contribution of the next higher order magneto-elastic terms, which are not included in Eq. \eqref{Eq1}~\cite{Linnik_2011}.

To conclude, we have proposed a concept for a nanoscale source of microwave magnetic field localized within $\sim$10~nm which shows high amplitude and narrow spectral width corresponding to a quality factor higher than 60. The physics of the proposed concept is based on driving the magnetization precession by coherent phonons in a laterally patterned ferromagnetic metal film. At resonance, the magnetization precession with an amplitude up to 1\% of the saturation magnetization produces an oscillating induction of $\sim$10 mT, much higher than available by existing microwave sources. The most important feature of the proposed concept is the ability to concentrate the strong ac-magnetic field on the nanoscale. Varying the pattern parameters of the ferromagnetic film should allow one to realize control of the ac-field in neighboring elements of the nanostructure by exciting different localized phonon modes. Driving by coherent phonons may be isolated from thermal effects by exciting the SAW through a bulk hypersonic wave approaching the nanostructure from the substrate~\cite{Bruggemann_2012}. The high sensitivity of the precession amplitude to the strength and orientation of the external magnetic field near the resonance condition allows one to tune the magnetization efficiently in and out of the magneto-phonon resonance. This opens a new way to manipulate magnetic states in nanoobjects. One can imagine a nanoscale chip, which combines a femtosecond semiconductor laser triggering the SAW that drives the magnetization precession in order to control single spin nanodevices located in the grooves of the patterned film, as shown in Fig. 1a. Specific examples are: nitrogen-vacancy spin states in diamond nanocrystals, spin molecules, and colloidal quantum dots. Their incorporation into the grooves of the SAW-driven ferromagnetic nanograting will allow control of their spin states by the ac-magnetic field. This provides attractive and exciting opportunities for further research and applications.

This work was supported by the Russian Scientific Foundation [grant number 16-12-10485] through support for experimental studies and nanolithography of the gratings; and by the Engineering and Physical Sciences Research Council [grant number EP/H003487/1] through support for growth and characterization of Galfenol films. The Volkswagen foundation [Grant number 90418] supported the theoretical work. The collaboration with TU Dortmund including working visits of A.P.D. to the Ioffe Institute was supported by the Deutsche Forschungsgemeinschaft in the frame of Collaborative Research Center TRR 160 (project B6). A.V.A. acknowledges the Alexander von Humboldt Foundation. We are grateful to Alexandra Kalashnikova and Leonid Golub for fruitful discussions and to Roman Dubrovin for AFM imaging.

\end{document}